**Solid-State Electrochemical Thermal Transistors with Large Thermal Conductivity Switching Widths**

*Zhiping Bian, Mitsuki Yoshimura, Ahrong Jeong, Haobo Li, Takashi Endo, Yasutaka Matsuo, Yusaku Magari, Hidekazu Tanaka, Hiromichi Ohta**

Z. Bian, M. Yoshimura
Graduate School of Information Science and Technology, Hokkaido University, N14W9, Kita, Sapporo 060-0814, Japan

A. Jeong, T. Endo, Y. Matsuo, Y. Magari, H. Ohta
Research Institute for Electronic Science, Hokkaido University, N20W10, Kita, Sapporo 001-0020, Japan
E-mail: hiromichi.ohta@es.hokudai.ac.jp

H. Li, H. Tanaka
SANKEN, Osaka University, Mihogaoka 8-1, Ibaraki, Osaka 567-0047, Japan

Thermal transistors that switch the thermal conductivity ($\kappa$) of the active layers are attracting increasing attention as thermal management devices. For electrochemical thermal transistors, several transition metal oxides (TMOs) have been proposed as active layers. After electrochemical redox treatment, the crystal structure of the TMO is modulated, which results in the $\kappa$ switching. However, the $\kappa$ switching width is still small (< 4 W m$^{-1}$ K$^{-1}$). In this study, we demonstrate that $LaNiO_x$-based solid-state electrochemical thermal transistors have a $\kappa$ switching width of 4.3 W m$^{-1}$ K$^{-1}$. Fully oxidised $LaNiO_3$ (on state) has a $\kappa$ of 6.0 W m$^{-1}$ K$^{-1}$ due to the large contribution of electron thermal conductivity ($\kappa_{ele}$, 3.1 W m$^{-1}$ K$^{-1}$). In contrast, reduced $LaNiO_{2.72}$ (off state) has a $\kappa$ of 1.7 W m$^{-1}$ K$^{-1}$ because the phonons are scattered by the oxygen vacancies. The $LaNiO_x$-based electrochemical thermal transistor exhibits excellent cyclability of $\kappa$ and the crystalline lattice of $LaNiO_x$. This electrochemical thermal transistor may be a promising platform for next-generation devices such as thermal displays.

## 1. Introduction

Reuse of waste heat resulting from the low conversion rate of primary energy is crucial for sustainable development. Low- to medium-temperature (100−300 °C) waste heat is the most difficult to reuse; the temperature is too low to generate jet steam for power generation. Although thermoelectric energy conversion technology is a solution, its efficiency is too low in this temperature range in air[1-4]. Thermal management technologies[5] such as thermal diodes[6-8] and thermal transistors[9-16] have recently attracted attention. Thermal diodes rectify the heat flow; thermal transistors electrically switch the heat flow on and off. We expect that thermal displays that visualise heat contrast using infrared cameras can be realised using thermal transistors. Thus, thermal transistors may be useful for reuse of waste heat.

For this purpose, electrical control of thermal conductivity ($\kappa$) in the active materials of thermal transistors is paramount. It necessitates a switch between the on state (high $\kappa$) and off state (low $\kappa$). Electrochemical[10-14] and electrostatic[15, 16] approaches offer pathways to govern the $\kappa$ of active materials. Although electrostatic methods provide rapid $\kappa$ control, their suitability for thermal display applications is limited by the requirement of an extremely thin active material around the heterointerface between the gate dielectric and the active material. We focus on electrochemical methods because they control the $\kappa$ of entire materials. Many studies have used ionic liquids such as organic electrolytes and water for electrochemical modulation of materials[10-12, 14]. However, this method is incompatible with integrated circuits, limiting its application. In our pursuit of thermal displays with substantial thermal conductance differences between the on and off states, we used all-solid-state electrochemical thermal transistors[13, 17, 18].

All-solid-state electrochemical thermal transistors, a cornerstone of advanced thermal management, harness the redox modulation of transition metal oxides (TMOs). Among many candidates[9-16], TMOs have emerged as promising materials. When subjected to electrochemical redox treatment by inserting and extracting metal ions[10] and oxide ions[12-14], TMOs undergo structural transformations leading to a switch in their $\kappa$ width. Despite these advancements, the challenge lies in achieving a substantial $\kappa$ switching width, a critical factor for practical application that is often limited (< 4 W $m^{-1}$ $K^{-1}$)[10, 12-14, 19].

To overcome this limitation, we chose $LaNiO_3$ as the active material for solid-state electrochemical thermal transistors. Bulk $LaNiO_3$ has comparably high electrical and thermal conductivity[20], indicating its potential for thermal conductivity modulation (**Supplementary Information S1**). The schematic depicted in **Fig. 1** introduces $LaNiO_x$ as the active layer, offering a wide $\kappa$ switching width. In the on state, $LaNiO_3$ has heightened electrical conductivity when fully oxidised, resulting in a significant contribution from the electron thermal conductivity. Conversely, the off state achieved through electrochemical reduction leads to oxygen vacancies, reduced electrical conductivity, and negligible electron thermal conductivity. Scattering of phonons by these vacancies manifests as a low thermal conductivity.

This study focuses on use of $LaNiO_x$-based electrochemical thermal transistors to address the limitations in $\kappa$ switching width. Our investigations indicated a $\kappa$ switching width of 4.3 W $m^{-1}$ $K^{-1}$. Fully oxidised $LaNiO_3$ (on state) exhibited a high $\kappa$ of 6.0 W $m^{-1}$ $K^{-1}$, primarily attributed to the contribution of electron thermal conductivity (3.1 W $m^{-1}$ $K^{-1}$). Conversely, reduced $LaNiO_{2.72}$ (off state) had a low $\kappa$ of 1.7 W $m^{-1}$ $K^{-1}$ due to scattering of phonons by oxygen vacancies. Both the reduction and oxidation processes exhibited a nearly linear change in thermal conductivity.

Furthermore, our study investigates the cyclability of $\kappa$ and the crystalline lattice of $LaNiO_x$-based thermal transistors. The exceptional performance of these electrochemical thermal transistors positions them as viable candidates for integration into next-generation devices, particularly thermal displays.

## 2. Results and Discussion

### 2.1. Electrochemical Thermal Transistor Fabrication and Operation

**Figure 2a** shows a schematic of the thermal transistor device structure, which is similar to those of our previous thermal transistors[13, 17, 18]. In this study, we inserted an extremely thin $SrCoO_x$ layer between the $LaNiO_3$ and solid electrolyte (Gd-doped $CeO_2$/YSZ) (**Supplementary Information S2**). As shown in **Supplementary Fig. S2,** when $LaNiO_3$ was grown on the $SrCoO_x$/GDC-buffered (001) YSZ, the crystallographic orientation was stronger than that without a buffer layer. X-ray reciprocal space mapping (RSM) around 113 YSZ diffraction spots (data not shown) confirmed that $LaNiO_3$ grown on $SrCoO_x$/GDC-buffered

(001) YSZ demonstrated the highest quality. Moreover, there was a positive correlation between the quality of $LaNiO_3$ and its thermal conductivity (**Supplementary Table S3**). Thus, use of the thermal transistor structure shown in **Fig. 2a** is crucial.

The setup for operation of the $LaNiO_x$ thermal transistor is shown in **Fig. 2b**. Electrochemical redox treatment was performed at 280 °C in air by applying a constant current of ±10 μA. During the redox reaction, we controlled the flown electron density $Q = (I{\cdot}t)/(e{\cdot}V)$ through the current application time (**Figs. 2c and 2d**), with a step of $Q = 2 \times 10^{21}$ cm$^{-3}$ marked as A – K, where $I$ is the flown current, $t$ is the application time, $e$ is the electron charge, and $V$ is the volume of the $LaNiO_x$ layer in the thermal transistor. In this study, electrochemical redox treatments were performed according to Faraday's law of electrolysis.

Reduction: $LaNiO_3 + 0.56e^- \rightarrow LaNiO_{2.72} + 0.28O^{2-}$

Oxidation: $LaNiO_{2.72} + 0.28O^{2-} \rightarrow LaNiO_3 + 0.56e^-$

The conductivity of the reduced state of $LaNiO_{2.72}$ (9 S cm$^{-1}$) also confirmed its identity as $LaNiO_{2.72}$[21].

Electrochemical redox treatment was initiated by applying a negative current to reduce $LaNiO_3$ to $LaNiO_{2.72}$ (**Fig. 2c**). As $Q$ increased, the absolute value of the voltage increased. The slope decreased after $2 \times 10^{21}$ cm$^{-3}$ and slightly increased after $2 \times 10^{21}$ cm$^{-3}$. When the current became positive, $LaNiO_{2.72}$ was oxidised to $LaNiO_3$ (**Fig. 2d**). As $Q$ increased, fluctuations occurred at $2 \times 10^{21}$ cm$^{-3}$ and $8 \times 10^{21}$ cm$^{-3}$. Overall, there was still an increasing trend and an absolute value between 3 V and 5 V, with minimal variations. The absence of steps in the process curve indicates that there were no new thermodynamically stable phases.

**2.2. Crystalline Lattice and Thermal Conductivity Changes during Redox Treatment**

Redox treatment induced reversible changes in the crystalline lattice of $LaNiO_x$ step by step (reduction A → F, oxidation: F → K), as evidenced by the out-of-plane x-ray diffraction patterns (**Fig. 3a**). Slight shifts in the 002 diffraction peak indicated modulations in the crystal structure, with lattice expansion observed after reduction and shrinkage observed after oxidation. The change in the lattice parameter $c$ is shown in **Fig. 3b**.

**Figure 4** and **Supplementary Fig. S3** illustrate the significant changes in the $\kappa$ of the $LaNiO_x$ during redox treatment. Time-domain thermoreflectance (TDTR) decay curves show a decrease in $\kappa$ from 5.9 W m$^{-1}$ K$^{-1}$ to 1.8 W m$^{-1}$ K$^{-1}$ after reduction (A → F) and an increase from 1.8 W m$^{-1}$ K$^{-1}$ to 5.9 W m$^{-1}$ K$^{-1}$ after oxidation (F → K). Both the reduction and oxidation processes exhibited a nearly linear change in thermal conductivity. The slight variations in lattice constants between the oxidised and reduced states are attributed to the disappearance and generation of oxygen vacancies. Through phonon scattering, oxygen vacancies contribute to a decrease in the lattice thermal conductivity.

However, there is a significant contrast in electrical conductivity ($\sigma$) between the oxidised (4250 S cm$^{-1}$) and reduced (9 S cm$^{-1}$) states (**Supplementary Table S4**). We estimated the electron thermal conductivity ($\kappa_{ele}$) by assuming the Wiedemann–Franz law; $\kappa_{ele} = L \cdot \sigma \cdot T$, where $L$ is the Lorentz number (2.44 × 10$^{-8}$ W Ω K$^{-2}$), and $T$ is the absolute temperature. The $\kappa_{ele}$ of the oxidised state reached 3.1 W m$^{-1}$ K$^{-1}$. According to the principle that observable thermal conductivity is the sum of lattice thermal conductivity ($\kappa_{lat}$) and $\kappa_{ele}$[22], we estimated the $\kappa_{lat}$ of $LaNiO_3$ (on state) to be 2.9 W m$^{-1}$ K$^{-1}$. This value aligns with previously reported values for bulk $LaNiO_3$[20], confirming the consistency of our findings. This substantial difference enabled the $LaNiO_x$ thermal transistor to modulate its thermal conductivity over a wide range. Furthermore, the linear and reversible changes in thermal conductivity indicate the potential of the $LaNiO_x$-based thermal transistor for precise thermal modulation.

### 2.3. Cycle Properties of Thermal Transistor

As shown in **Fig. 5 and Supplementary Fig. S4**, the $LaNiO_x$-based thermal transistor exhibits exceptional cycling properties. The on state ($LaNiO_3$) has a higher average thermal conductivity (6.0 W m$^{-1}$ K$^{-1}$) than the off state ($LaNiO_{2.72}$, 1.7 W m$^{-1}$ K$^{-1}$). Seven cycles are shown in the figure. The TDTR decay of $LaNiO_3$ was faster than that of $LaNiO_{2.72}$ (**Fig. 5a**). The TDTR curves for each cycle overlapped significantly, indicating excellent repeatability.

### 2.4. Comparison with Other TMO Thermal Transistors

**Figure 6** compares the thermal conductivity switching widths of different TMOs, indicating the unparalleled performance of $LaNiO_x$-based thermal transistors, with a thermal conductivity switching width of approximately 4.3 W m$^{-1}$ K$^{-1}$, significantly greater than that of other TMOs. The wide switching range indicates the exceptional versatility of $LaNiO_x$ in modulating its thermal conductivity. The inherent ability of $LaNiO_x$-based thermal transistors to linearly transition between high thermal conductivity in the fully oxidised state and low thermal conductivity in the reduced state positions them at the forefront of TMOs for thermal management applications. The nuanced control over thermal conductivity and the stability demonstrated in the cycling properties (**Fig. 5**) underscore the potential of $LaNiO_x$-based thermal transistors for application in next-generation thermal displays and other advanced thermal management systems.

## 3. Conclusion

This study presents a breakthrough with $LaNiO_x$-based electrochemical thermal transistors, with an exceptional $\kappa$ switching width of 4.3 W m$^{-1}$ K$^{-1}$. The fully oxidised $LaNiO_3$ (on state) produced a $\kappa$ of 6.0 W m$^{-1}$ K$^{-1}$, primarily attributed to a substantial contribution from electron thermal conductivity (3.1 W m$^{-1}$ K$^{-1}$). In contrast, the reduced $LaNiO_{2.72}$ (off state) had a low $\kappa$ of 1.7 W m$^{-1}$ K$^{-1}$ due to negligible electron thermal conductivity (0.007 W m$^{-1}$ K$^{-1}$) and phonon scattering caused by oxygen vacancies. The $LaNiO_x$-based electrochemical thermal transistor demonstrated outstanding $\kappa$ cyclability while maintaining the structural integrity of the $LaNiO_x$ crystalline lattice, making it a promising candidate for integration into next-generation devices, particularly thermal displays.

## 4. Experimental Section

*Fabrication of thermal transistors*: $LaNiO_3$ films were heteroepitaxially grown on $SrCoO_x$/10%-Gd-doped $CeO_2$ (GDC)-buffered (001)-oriented YSZ substrates using pulsed laser deposition (PLD). First, approximately10-nm-thick GDC was heteroepitaxially grown on a YSZ (10 mm × 10 mm × 0.5 mm, double-sided polished, crystal base) substrate at 770 °C in an oxygen atmosphere (10 Pa). Approximately 2-nm-thick $SrCoO_x$ was grown on GDC in the same conditions. Focused KrF excimer laser pulses ($\lambda$ = 248 nm, fluence ~2 J cm$^{-2}$ pulse$^{-1}$, repetition rate = 10 Hz) were irradiated onto the ceramic target of GDC. Subsequently, an approximately 80-nm-thick $LaNiO_3$ film was heteroepitaxially grown on the GDC film at 625 °C in an oxygen atmosphere (25 Pa). The laser fluence was approximately 1.6 J cm$^{-2}$ pulse$^{-1}$. After film growth, the sample was cooled to room temperature in a PLD chamber in an oxygen atmosphere (25 Pa). An approximately 50-nm-thick Pt film was sputtered on the top surface of the $LaNiO_3$ epitaxial film, followed by an approximately 50-nm-thick Pt film sputtered on the backside of the YSZ substrate. Pt sputtering was performed at room temperature. The samples were cut into four squares (5 mm × 5 mm).

*Electrochemical redox treatment*: thermal transistor (5 mm × 5 mm) was placed on a Pt-coated glass substrate and heated at 280 °C in air. Electrochemical redox treatment was performed by applying a constant current of ±10 μA, after which the sample was immediately cooled to room temperature.

*Crystallographic analyses*: The crystalline phase, orientation, and lattice parameters of the resultant films were analysed using high-resolution x-ray diffraction (Cu K$\alpha_1$, $\lambda$ = 1.54059 Å, ATX-G, Rigaku). Out-of-plane Bragg diffraction patterns and reciprocal space mappings (RSMs) were measured at room temperature to clarify changes in the crystalline phase of $LaNiO_x$. The lattice parameters were calculated from the diffraction peaks. The atomic arrangements of the $LaNiO_3$ films were visualised using a STEM (JEM-ARM200CF, JEOL) operated at 200 keV.

*Measurement of electrical properties of $LaNiO_x$ layers*: To measure the electrical conductivity ($\sigma$) of the $LaNiO_x$ layers after redox treatment, we mechanically attached Au foil on the film surface while Pt films were deposited only on the backside of the YSZ substrate[13]. The $LaNiO_x$ films were oxidised and reduced electrochemically at 280 °C in air using the Au foil as the electrode. The $\sigma$ of the $LaNiO_3$ (on state) and $LaNiO_{2.72}$ (off state) films was measured using the DC four-probe method with a van der Pauw electrode configuration at room temperature in air.

*Thermal conductivity measurements*: The κ of the $LaNiO_3$ layers perpendicular to the substrate surface was measured through time-domain thermoreflectance (TDTR, PicoTR, PicoTherm). The top Pt film was used as the transducer. The decay curves of the TDTR signals were simulated to obtain κ. The specific heat capacities of the layers used for the TDTR simulation were Pt: 132 J $kg^{-1}$ $K^{-1}$ [23]; $LaNiO_3$: 448 J $kg^{-1}$ $K^{-1}$ [24], and YSZ: 460 J $kg^{-1}$ $K^{-1}$ [25]. Details of the TDTR method are described in our previous studies[13, 17, 26-28]. In treatment of the thermal conductivity values, as there were uncertainties such as the position of the baseline, position of time zero, and noise in the signal, we used error bars for ±15% of the obtained values.

**Acknowledgements**

This research was supported by Grants-in-Aid for Scientific Research A (22H00253) and Innovative Areas (19H05791) from the Japan Society for Promotion of Science (JSPS). Part of this work was supported by the Crossover Alliance to Create the Future with People, Intelligence and Materials, and by the Network Joint Research Centre for Materials and Devices. Part of this work was supported by the Advanced Research Infrastructure for Materials and Nanotechnology, Japan (Grant Number JPMXP1222UT0055) and the Ministry of Education, Culture, Sports, Science, and Technology (MEXT). Z.B. was supported by a Hokkaido University DX Doctoral Fellowship (JPMJSP2119).

Received: ((will be filled in by the editorial staff))
Revised: ((will be filled in by the editorial staff))
Published online: ((will be filled in by the editorial staff))

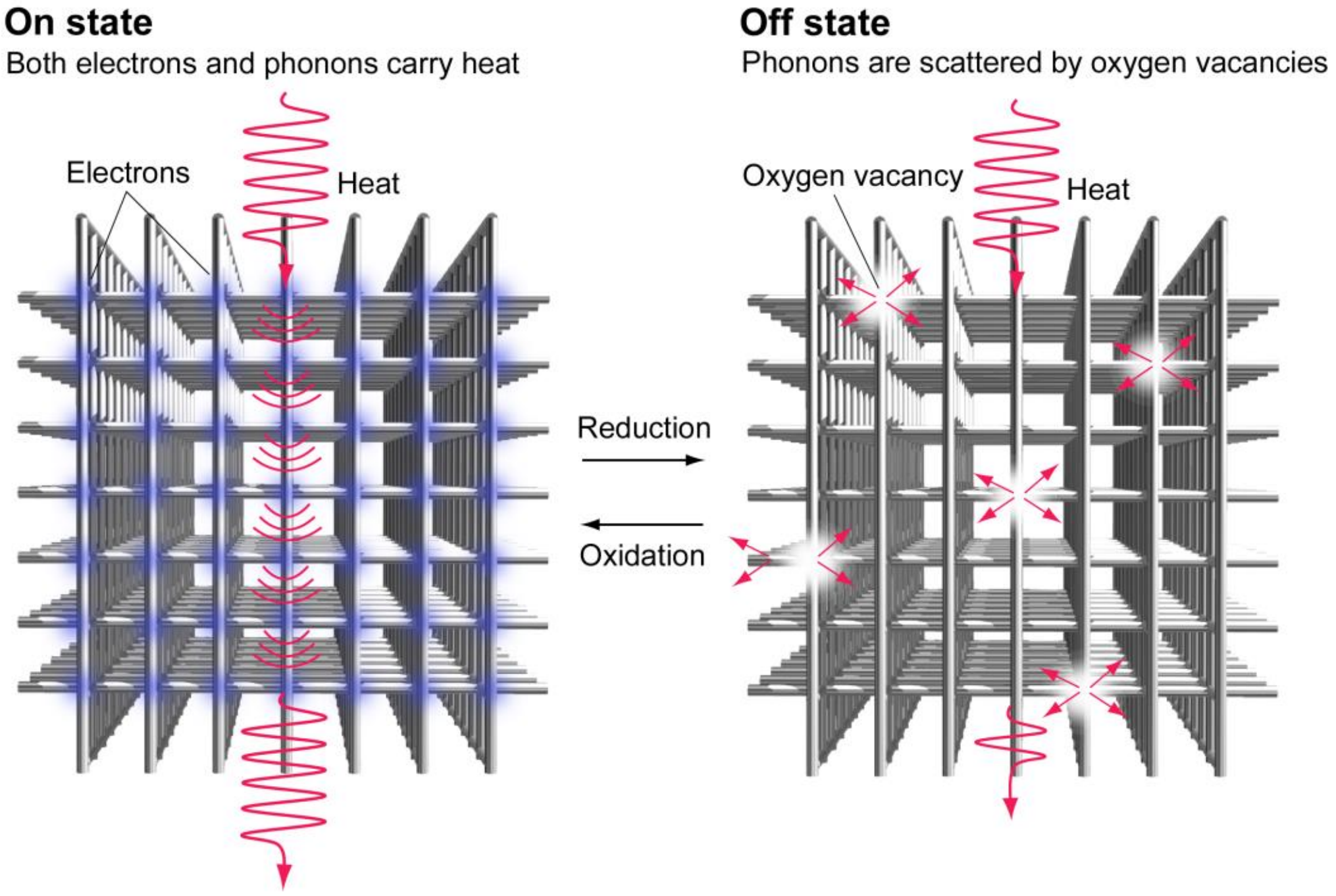

**Figure 1. Strategy of electrochemical thermal transistors with transition metal oxide active layer with large thermal conductivity switching width.** (Left) Diagram of on state. Fully oxidised TMOs show high electrical conductivity. Both electrons and phonons carry heat. The thermal transistor shows high thermal conductivity. (Right) Diagram of off state. Electrochemical reduction treatment produces oxygen vacancies. The reduced TMOs show low electrical conductivity. The electron thermal conductivity is negligible. Phonons are scattered by oxygen vacancies. The thermal transistor shows low thermal conductivity.

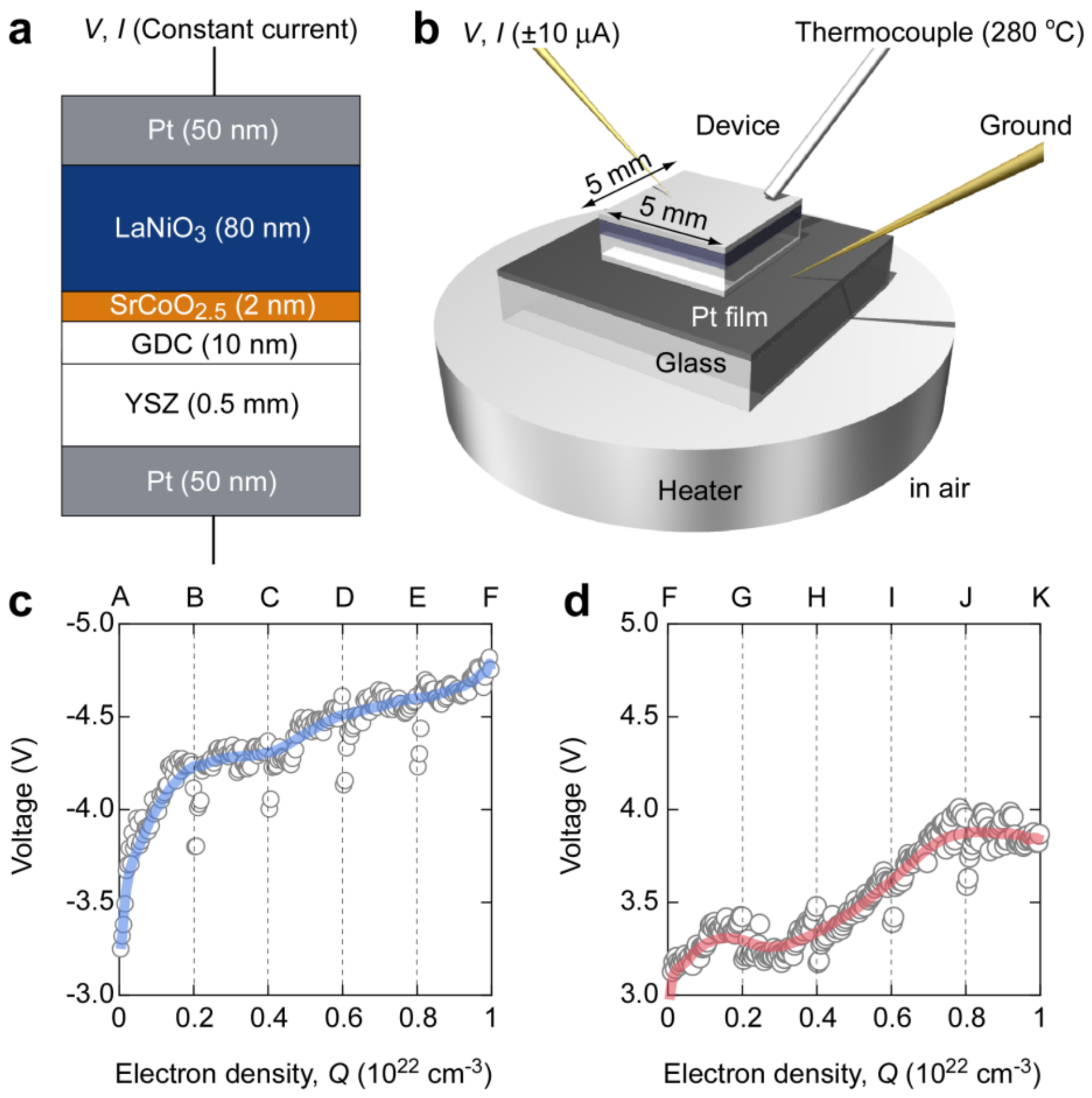

**Figure 2. LaNiO$_x$-based thermal transistor operation.** (a) Structure of thermal transistor composed of six layers: 50-nm-thick Pt film, 80-nm-thick LaNiO$_x$ film, 2-nm-thick SrCoO$_x$ film, 10-nm-thick Gd-doped $CeO_2$ (GDC) film, 0.5-mm-thick (001) YSZ single-crystal substrate, and 40-nm-thick Pt film. (b) Schematic of LaNiO$_x$ thermal transistor. The transistor was placed on a Pt-coated glass substrate and heated at 280 °C in air. A K-type thermocouple was used to monitor the transistor surface temperature. The transistor measured 5 mm × 5 mm. A constant negative current (−10 μA) was applied for reduction; a constant positive current (+10 μA) was applied for oxidation. (c)(d) Changes in observed DC voltage of thermal transistor during (c) reduction from $LaNiO_3$ to $LaNiO_{2.72}$ and (d) oxidation from $LaNiO_{2.72}$ to $LaNiO_3$ with a step of $Q = 2 \times 10^{21}$ cm$^{-3}$.

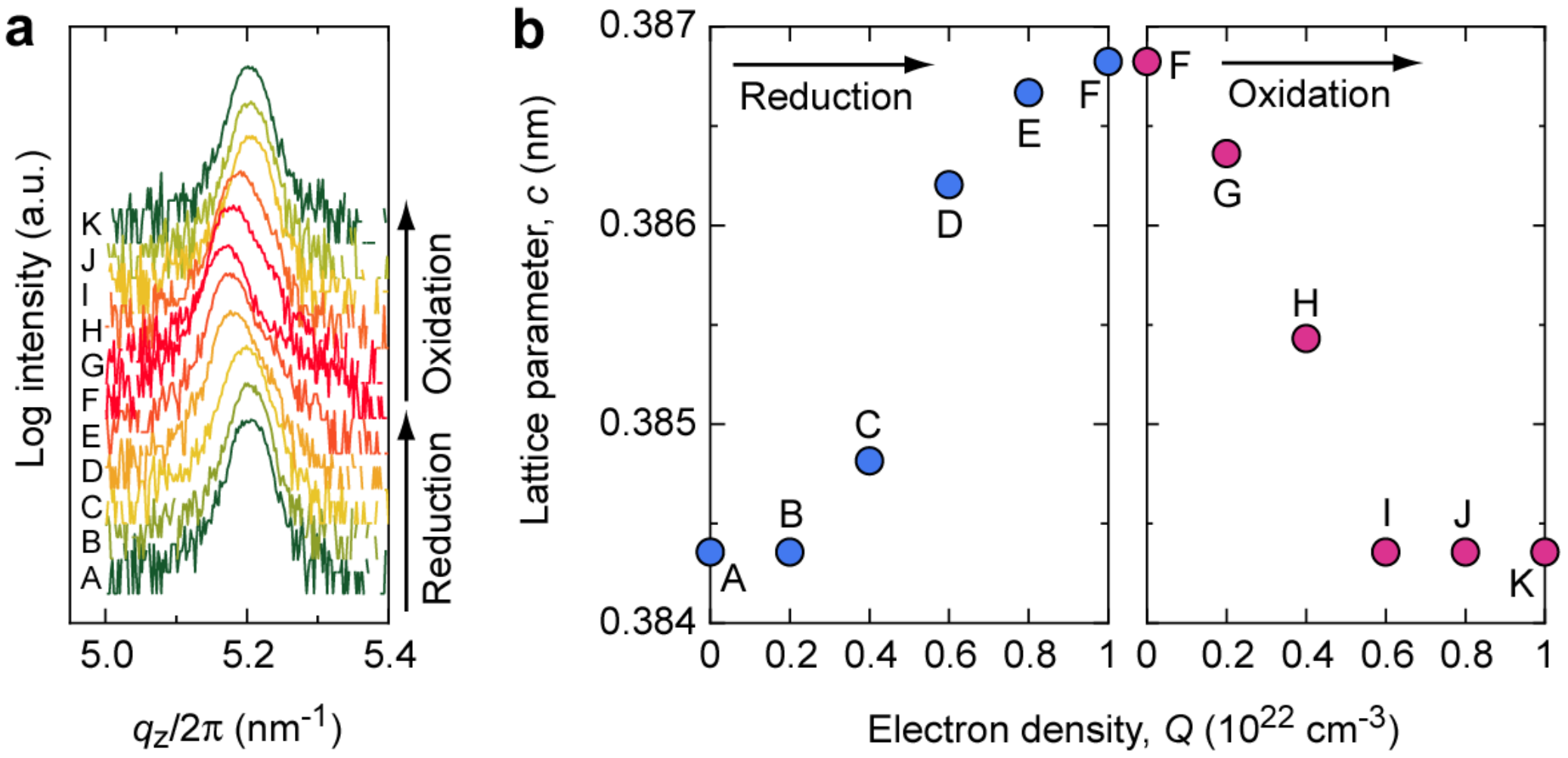

**Figure 3. Change in crystalline lattice of $LaNiO_x$ layer after redox treatment.** (a) Change in out-of-plane XRD patterns after redox treatment with a step of $Q = 2 \times 10^{21}$ $cm^{-3}$. The 002 diffraction peak shifted to a smaller $q_z$ side after reduction and a larger $q_z$ side after oxidation. These shifts were entirely reversible. (b) Changes in lattice parameter $c$ of $LaNiO_x$ as a function of $Q$. A lattice expansion of approximately 0.6% occurred after reduction.

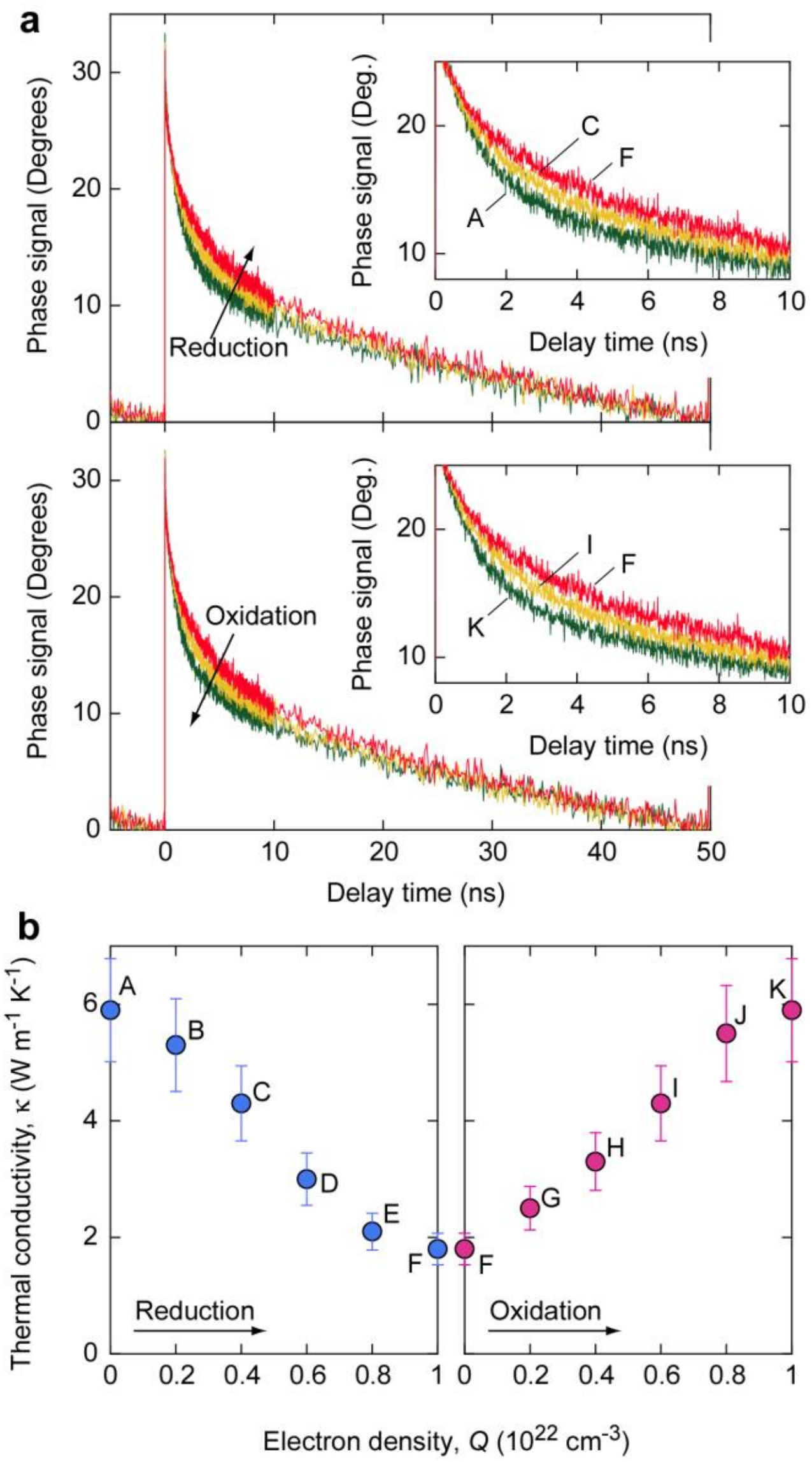

**Figure 4. Change in thermal conductivity (*κ*) of $LaNiO_x$ layer after redox treatment.** (a) TDTR decay curves of thermal transistor after (upper) reduction and (lower) oxidation treatment. (b) Changes in thermal conductivity of $LaNiO_x$ layer after (left) reduction and (right) oxidation treatment with a step of $Q = 2 \times 10^{21}$ $cm^{-3}$. The $\kappa$ of $LaNiO_3$ was 5.9 W $m^{-1}$ $K^{-1}$; it decreased almost linearly with $Q$ after reduction. After oxidation, the $\kappa$ increased almost linearly with $Q$ and returned to 5.9 W $m^{-1}$ $K^{-1}$.

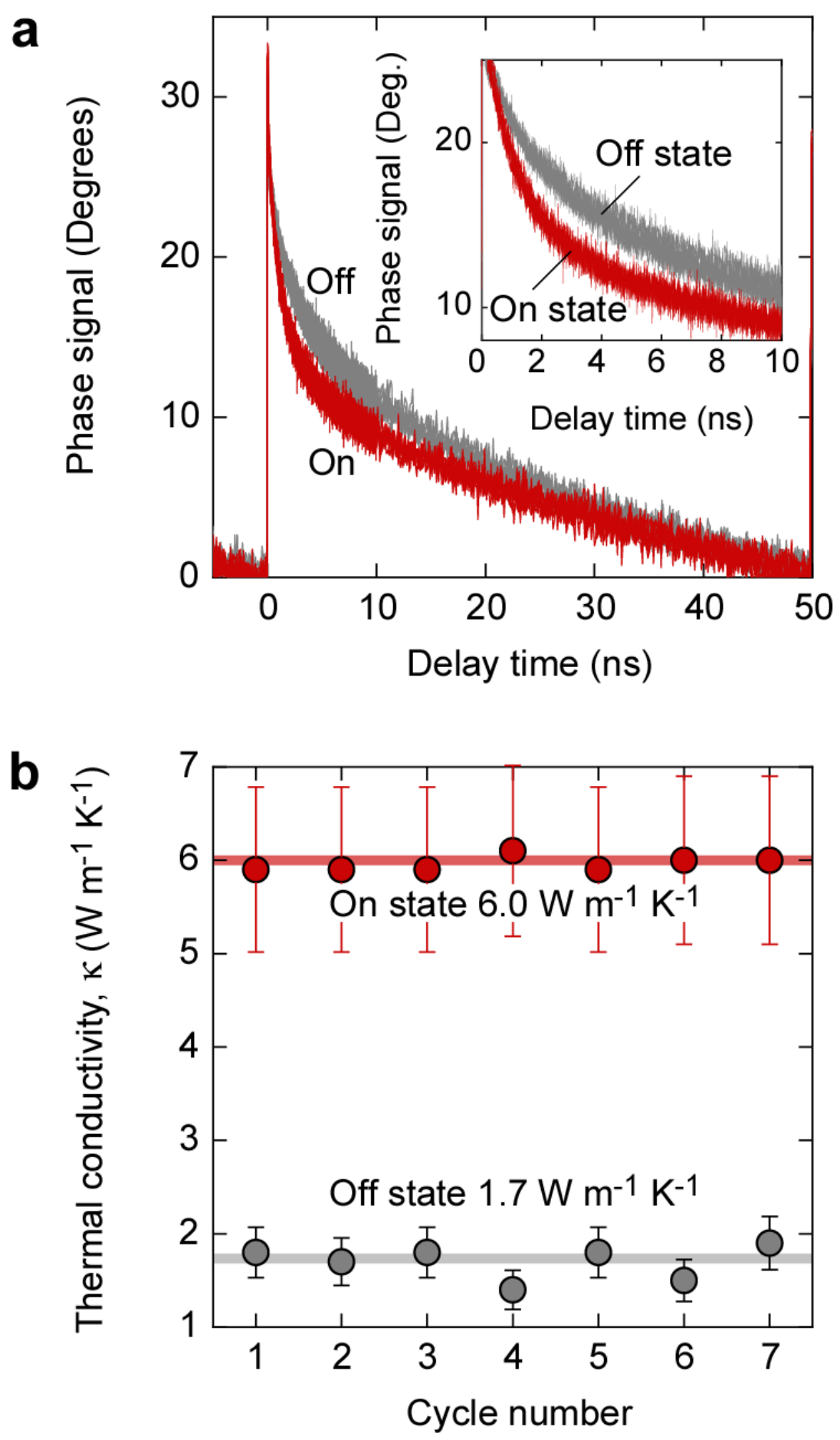

**Figure 5. Cycle properties of LaNiO$_x$-based thermal transistor.** (a) Changes in TDTR decay curves (seven cycles overlapped). The TDTR decay of the $LaNiO_3$ layer was faster than that of the $LaNiO_{2.72}$ layer. (b) Change in thermal conductivities of $LaNiO_3$ and $LaNiO_{2.72}$ layers after redox cycling. The average thermal conductivities of the $LaNiO_3$ (on state) layer and $LaNiO_{2.72}$ (off state) layer were 6.0 W $m^{-1}$ $K^{-1}$ and 1.7 W $m^{-1}$ $K^{-1}$, respectively. The electron thermal conductivity of the $LaNiO_3$ (on state) layer was high (3.1 W $m^{-1}$ $K^{-1}$). The electron thermal conductivity of the $LaNiO_{2.72}$ (off state) layer was negligible.

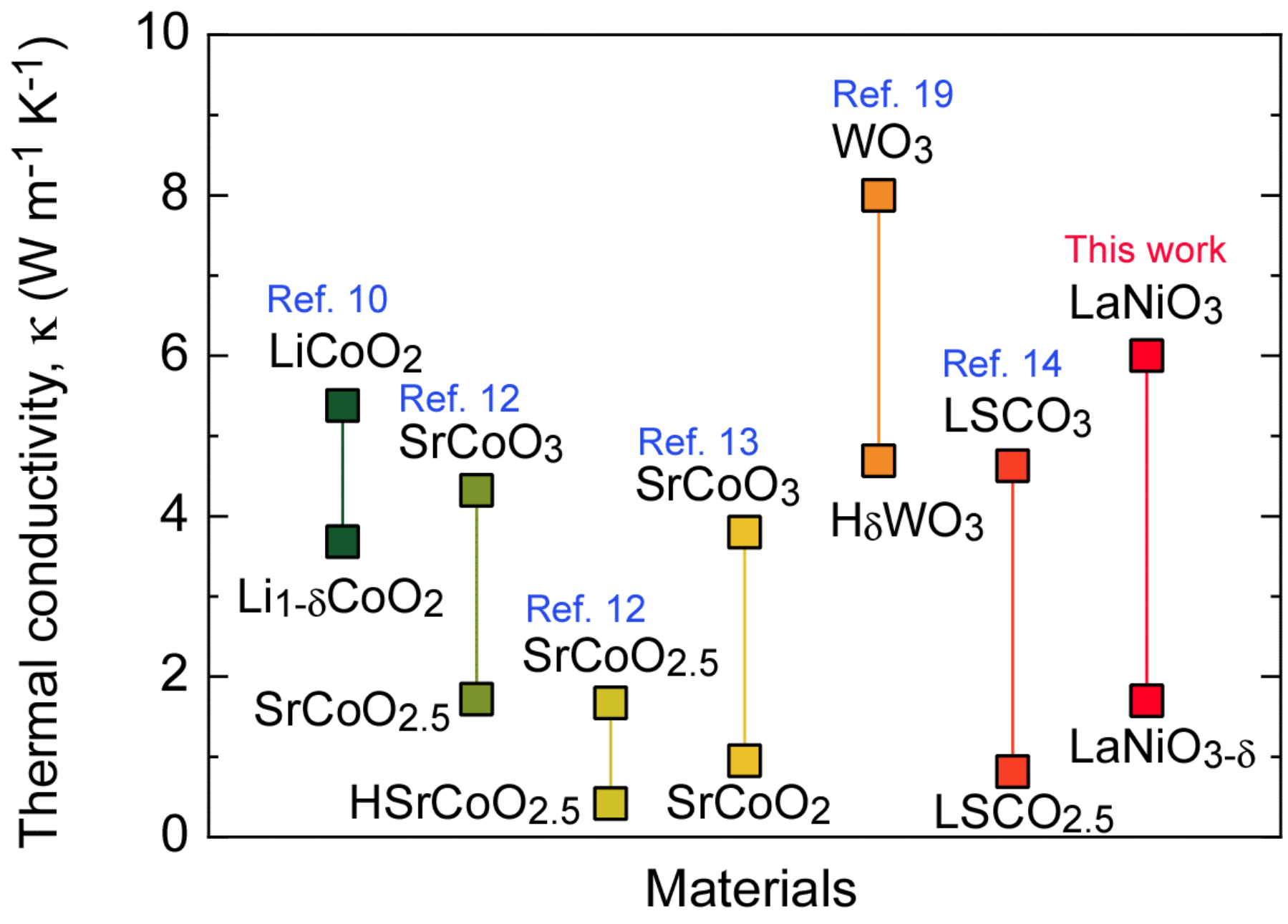

**Figure 6. Comparison of thermal conductivity switching widths of several transition metal oxides.** $LaNiO_x$-based thermal transistors exhibited large thermal conductivity switching widths (~4.3 W $m^{-1}$ $K^{-1}$). Data for $LiCoO_2$ ↔ $Li_{1-\delta}CoO_2$ are from Ref. 10[10], $SrCoO_3$ ↔ $SrCoO_{2.5}$ and $SrCoO_{2.5}$ ↔ $HSrCoO_{2.5}$ are from Ref. 12[12], $SrCoO_3$ ↔ $SrCoO_2$ are from Ref. 13[13], $WO_3$ ↔ $H_{\delta}WO_3$ are from Ref. 19[19], and $La_{0.5}Sr_{0.5}CoO_3$ (LSCO3) ↔ $La_{0.5}Sr_{0.5}CoO_{2.5}$ ($LSCO_{2.5}$) are from Ref. 14[14].

The table of contents entry should be 50–60 words long and should be written in the present tense. The text should be different from the abstract text.

Z. Bian, M. Yoshimura, A. Jeong, H. Li, T. Endo, Y. Matsuo, Y. Magari, H. Tanaka, H. Ohta*

**Solid-State Electrochemical Thermal Transistors with Large Thermal Conductivity Switching Widths**

ToC figure ((Please choose one size: 55 mm broad × 50 mm high **or** 110 mm broad × 20 mm high.  Please do not use any other dimensions))

# Supporting Information

**Solid-State Electrochemical Thermal Transistors with Large Thermal Conductivity Switching Widths**

*Zhiping Bian, Mitsuki Yoshimura, Ahrong Jeong, Haobo Li, Takashi Endo, Yasutaka Matsuo, Yusaku Magari, Hidekazu Tanaka, Hiromichi Ohta**

## S1. Selection of $LaNiO_3$ as the active layer

In this study, we focused on perovskite-structured $Ln$NiO$_3$ ($Ln$ = La, Nd, and Sm) as the active layer for the solid-state electrochemical thermal transistor. **Table S1** summarizes the electrical conductivity ($\sigma$) and the thermal conductivity ($\kappa$) of bulk $Ln$NiO$_3$ ($Ln$ = La, Nd, and Sm) at room temperature[20]. At room temperature, bulk $Ln$NiO$_3$ ($Ln$ = La, Nd, and Sm) shows the following $\kappa$; $LaNiO_3$: 10.7 W m$^{-1}$ K$^{-1}$, $NdNiO_3$: 6.5 W m$^{-1}$ K$^{-1}$, and $SmNiO_3$: 4.0 W m$^{-1}$ K$^{-1}$. Thus, $LaNiO_3$ is a promising candidate as the active layer of the thermal transistors. It should be noted that the $\sigma$ of $LaNiO_3$ is 10500 S cm$^{-1}$. We assumed the Wiedemann-Frantz law for the estimation of electron contribution to the observed thermal conductivity ($\kappa_{ele} = L\cdot\sigma\cdot T$, where $L$ is the Lorentz number of $2.44 \times 10^{-8}$ W Ω K$^{-2}$ and $T$ is the absolute temperature of 298 K) and obtained $\kappa_{ele}$ of 7.6 W m$^{-1}$ K$^{-1}$ for $LaNiO_3$. This reflects the lattice thermal conductivity ($\kappa_{lat}$) of $LaNiO_3$ is 3.1 W m$^{-1}$ K$^{-1}$.

**Table S1**. **The electrical conductivity ($\sigma$) and the thermal conductivity ($\kappa$) of bulk $Ln$NiO$_3$ ($Ln$ = La, Nd, and Sm) at room temperature.** The ionic radius data is from Shannon's report.[2]

| J.-S. Zhou *et al.*, *PRB* **67**, 020404(R) (2003)[1] | $LaNiO_3$ | $NdNiO_3$ | $SmNiO_3$ |
|---|---|---|---|
| Ionic radius of $Ln^{3+}$ ion (Å) (C.N. = 12) | 1.36 | 1.27 | 1.24 |
| Electrical conductivity at RT (S cm$^{-1}$) | 10500 | 3400 | --- |
| Total thermal conductivity, $\kappa$ (W m$^{-1}$ K$^{-1}$) | 10.7 | 6.5 | 4.0 |
| Electron thermal conductivity, $\kappa_{ele}$ (W m$^{-1}$ K$^{-1}$) | 7.6 | 2.5 | 0 |
| Lattice thermal conductivity, $\kappa_{lat}$ (W m$^{-1}$ K$^{-1}$) | 3.1 | 4.0 | 4.0 |

To check the potential of $LaNiO_3$ epitaxial films as the active layer of the thermal transistors, we fabricated $LaNiO_3$ epitaxial films on (001) $SrTiO_3$ substrates and measured the electrical and thermal conductivity of the resultant films at room temperature (**Table S2**). The $\sigma$ of the resultant $LaNiO_3$ film was only 135 S cm$^{-1}$, two orders of magnitude smaller than that of bulk. The $\kappa$ in the out-of-plane of the $LaNiO_3$ film was 7.3 W m$^{-1}$ K$^{-1}$. If we assumed the Wiedemann-Frantz law for the estimation of $\kappa_{ele}$, the $\kappa_{ele}$ of the $LaNiO_3$ film was only 0.1 W m$^{-1}$ K$^{-1}$, reflecting the $\kappa_{lat}$ of $LaNiO_3$ is 7.2 W m$^{-1}$ K$^{-1}$.

**Table S2**. **The electrical and thermal conductivity of the $LaNiO_3$ epitaxial films on (001) $SrTiO_3$ at room temperature.**

| This study | $LaNiO_3$ |
|---|---|
| Ionic radius of $Ln^{3+}$ ion (Å) (C.N. = 12) | 1.36 |
| Electrical conductivity at RT (S cm$^{-1}$) | 135 |
| Total thermal conductivity, $\kappa$ (W m$^{-1}$ K$^{-1}$) | 7.3 |
| Electron thermal conductivity, $\kappa_{ele}$ (W m$^{-1}$ K$^{-1}$) | 0.1 |
| Lattice thermal conductivity, $\kappa_{lat}$ (W m$^{-1}$ K$^{-1}$) | 7.2 |

Since the estimated $\kappa_{lat}$ of the $LaNiO_3$ film on (001) $SrTiO_3$ substrate is higher than that of the bulk, there is a possibility that we underestimated the $\kappa_{ele}$ of the $LaNiO_3$ film. To clarify the origin of it, we observed the microstructure using Cs-corrected scanning transmission electron microscopy (**Fig. S1**). The columnar structure is visualized in the $LaNiO_3$ film (**Fig. S1a**). The magnified image (**Fig. S1b**) reveals that there are many planer defects due to the formation of Ruddlesden-Popper phases. Since the electron transport is suppressed by the planer defects, the observed electrical conductivity in the in-plane direction would be lower than that in the out-of-plane direction.

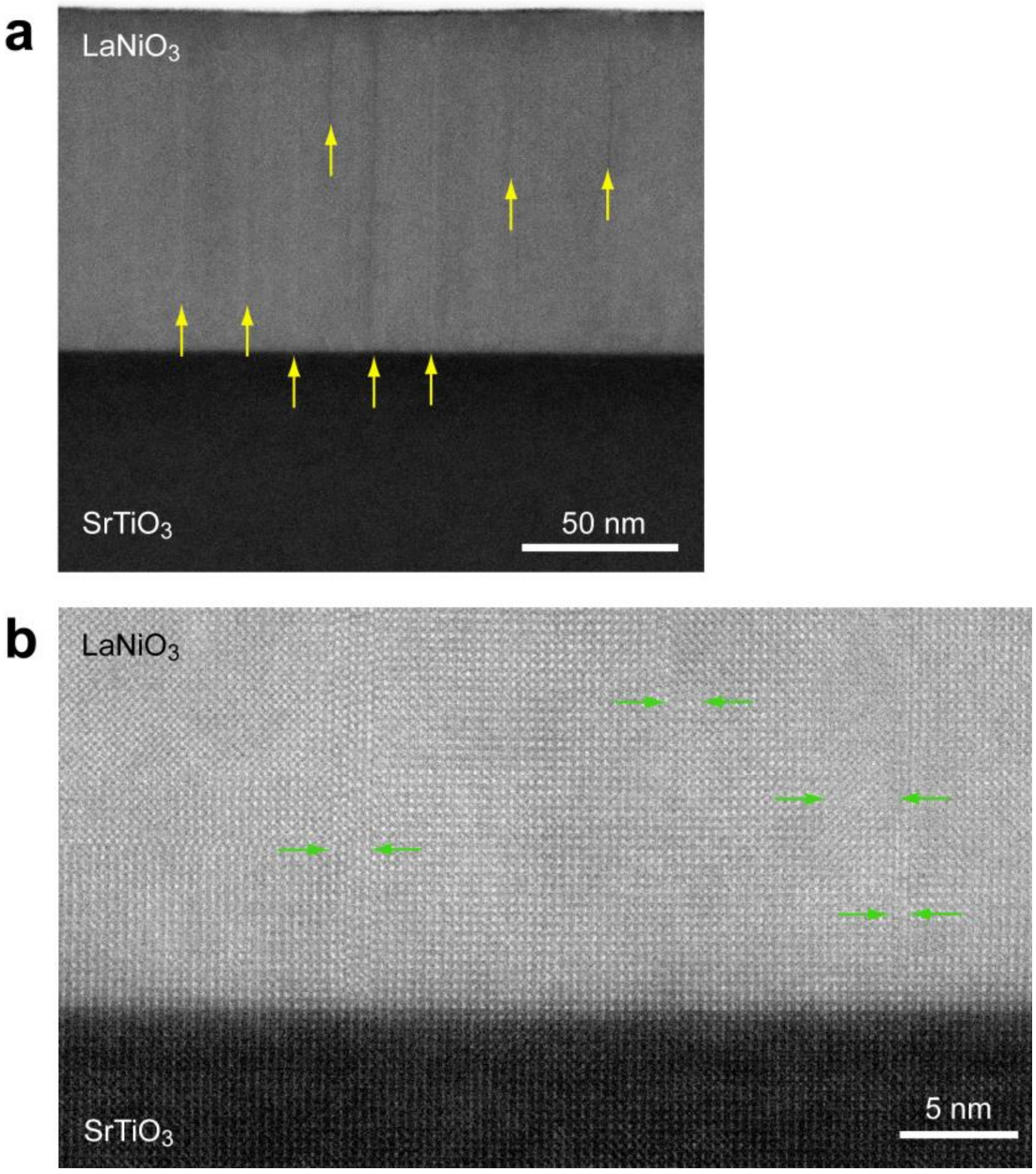

**Figure S1. Microstructure of the $LaNiO_3$ film grown on (001) $SrTiO_3$ substrate. a,** Low magnification cross-sectional HAADF-STEM image. Stripe patterns (yellow arrows) indicates that columnar growth of $LaNiO_3$ occurred. **b,** Lattice image around the $LaNiO_3$/$SrTO_3$ heterointerface. Planar defects (green arrows) are visualized.

## S2. Fabrication of $LaNiO_3$-based solid-state thermal transistors

Firstly, we fabricated $LaNiO_3$ films directly on (001) YSZ substrates. **Figure S2a** shows the out-of-plane XRD pattern of the resultant film. Together with 001 and 002 diffraction peaks of $LaNiO_3$, 110 diffraction peaks of $LaNiO_3$ are seen with 002 YSZ, indicating the mixed orientation of the film. The out-of-plane rocking curve of the 002 $LaNiO_3$ (**Fig. S2d**) is broad (the full width at half maximum, FWHM ~3.1°). Then, we fabricated $LaNiO_3$ films on GDC-buffered (001) YSZ substrates. As shown in **Fig. S2b**, intense diffraction peaks of 001 and 002 $LaNiO_3$ are seen together with 002 GDC and 002 YSZ. The FWHM of the 002 $LaNiO_3$ is 1.6° as shown in **Fig. 2e**. The reciprocal space mapping (RSM, not shown) of the film revealed that the $LaNiO_3$ is heteroepitaxially grown on the GDC-buffered YSZ substrate. To further improve the crystallographic orientation of the film, we inserted thin (~2 nm) $SrCoO_x$ layer between the $LaNiO_3$ and GDC-buffered substrate. The out-of-plane XRD peaks became stronger (**Fig. S2c**) and the tilting became small (1.2°) as shown in **Fig. S2f**.

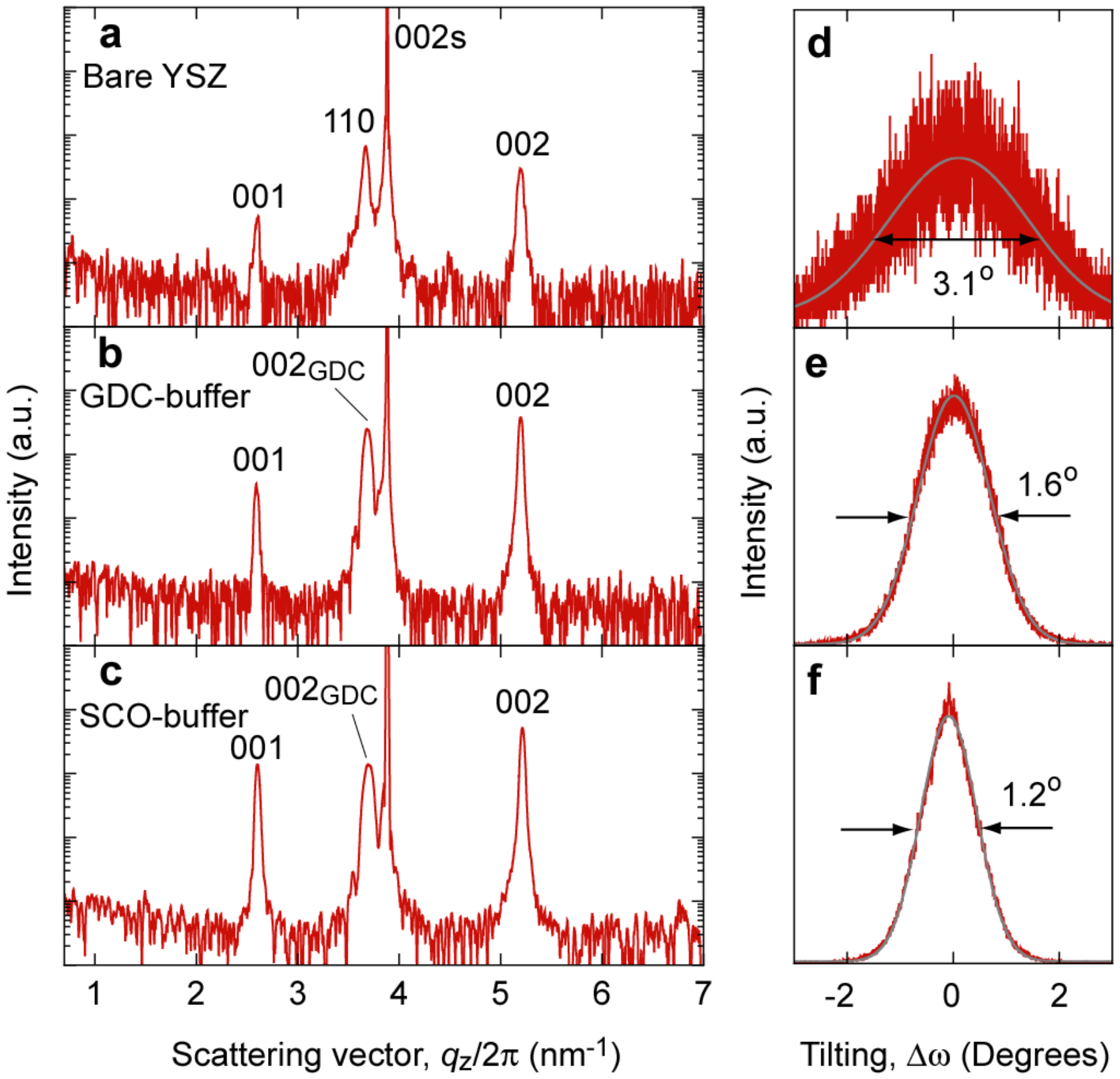

**Figure S2. XRD patterns of the $LaNiO_3$ films grown on various substrates. a, b, c,** Out-of-plane Bragg diffraction patterns of the $LaNiO_3$ films grown on (a) bare (001) YSZ substrate, (b) GDC-buffered (001) YSZ substrate, and (c) $SrCoO_x$/GDC-buffered (001) YSZ substrate. **d, e, f,** Out-of-plane rocking curves of 002 $LaNiO_3$ of the films on (d) bare (001) YSZ substrate, (e) GDC-buffered (001) YSZ substrate, and (f) $SrCoO_x$/GDC-buffered (001) YSZ substrate.

Then, we measured the $\sigma$ and the $\kappa$ of the resultant $LaNiO_3$ films on the various substrates at room temperature. **Table S3** summarizes the results. The $LaNiO_3$ film on the $SrCoO_x$-buffered substrate showed the highest $\sigma$ of 4200 S $cm^{-1}$, reflecting the improvement of the crystallographic orientation of the $LaNiO_3$ film. The out-of-plane $\kappa$ of the $LaNiO_3$ film on the $SrCoO_x$-buffered substrate was the highest (5.7 W $m^{-1}$ $K^{-1}$) among the films on different three substrates. We estimated the $\kappa_{ele}$ and the $\kappa_{lat}$ of the $LaNiO_3$ films. The $\kappa_{lat}$ of the $LaNiO_3$ film on $SrCoO_x$-buffered substrate was 2.7 W $m^{-1}$ $K^{-1}$, similar to that of bulk (3.1 W $m^{-1}$ $K^{-1}$).

**Table S3. Electrical and thermal conductivity of the $LaNiO_3$ films grown on the three different substrates.**

| | Bare YSZ | GDC-buffered | SCO-buffered |
|---|---|---|---|
| Electrical conductivity, $\sigma$ (S $cm^{-1}$) | 500 | 1200 | 4200 |
| Thermal conductivity, $\kappa$ (W $m^{-1}$ $K^{-1}$) | 1.6 | 4 | 5.7 |
| Electron thermal conductivity, $\kappa_{ele}$ (W $m^{-1}$ $K^{-1}$) | 0.36 | 0.88 | 3.0 |
| Lattice thermal conductivity, $\kappa_{lat}$ (W $m^{-1}$ $K^{-1}$) | 1.2 | 3.1 | 2.7 |

Then, we reduced the $LaNiO_3$ film on SCO-buffered (001) YSZ substrate electrochemically, and measured the $\sigma$ and the $\kappa$ (**Table S4**). The reduction treatment by applying total $Q$ of $1 \times 10^{22}$ $cm^{-3}$ results in the significant reduction of both $\sigma$ and the $\kappa$.

**Table S4. Electrical and thermal conductivity of the $LaNiO_3$ films on $SrCoO_x$/GDC-buffered substrate (oxidized state and reduced state).**

| | Oxidized | Reduced |
|---|---|---|
| Electrical conductivity, $\sigma$ (S $cm^{-1}$) | 4200 | 9 |
| Thermal conductivity, $\kappa$ (W $m^{-1}$ $K^{-1}$) | 5.9 | 1.8 |
| Electron thermal conductivity, $\kappa_{ele}$ (W $m^{-1}$ $K^{-1}$) | 3.1 | 0.006 |
| Lattice thermal conductivity, $\kappa_{lat}$ (W $m^{-1}$ $K^{-1}$) | 2.8 | 1.8 |

### S3. Repeated thermal conductivity measurements of the $LaNiO_3$-based solid-state thermal transistors

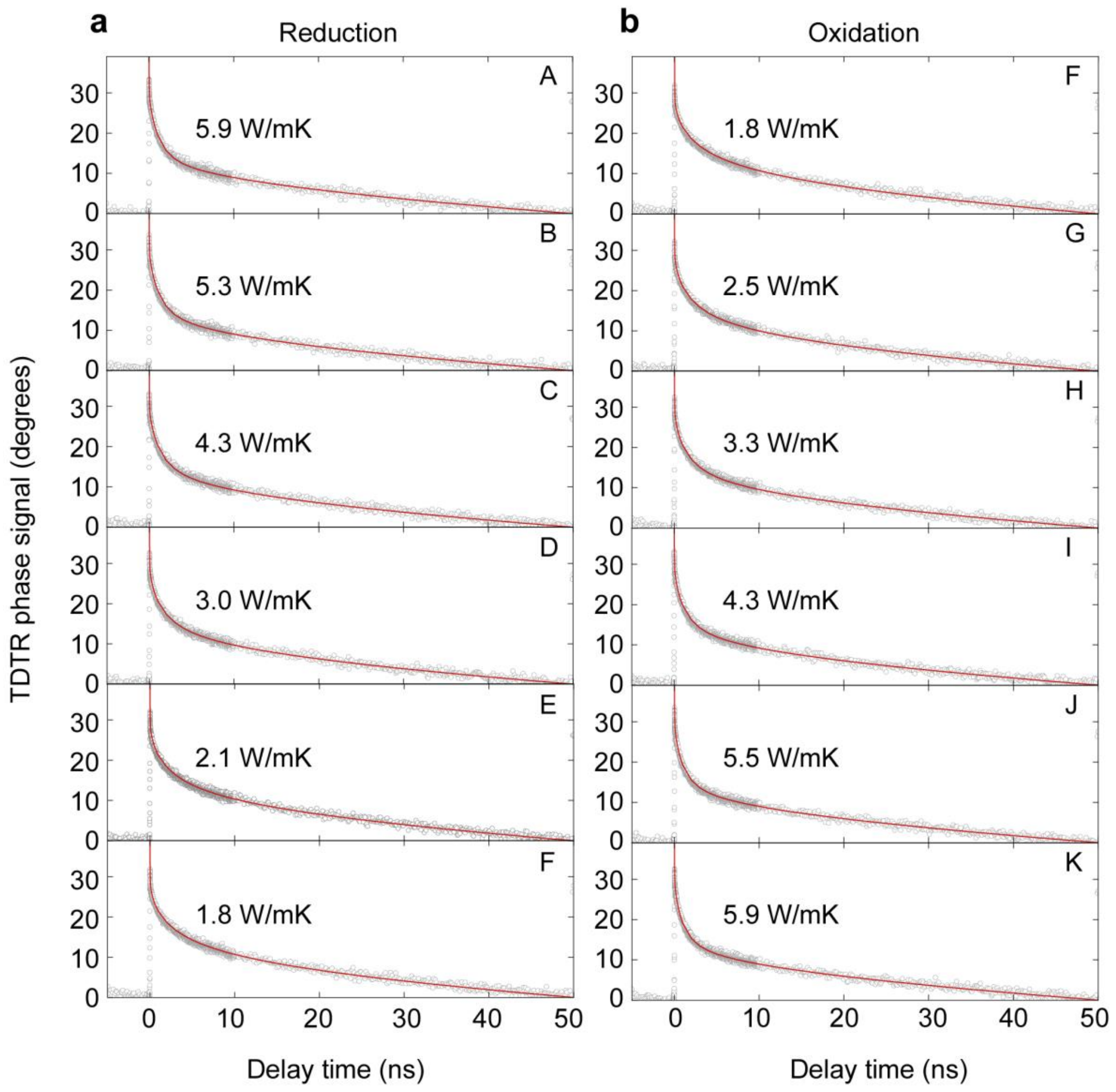

**Figure S3. Change in the thermal conductivity of the $LaNiO_3$ layer in the thermal transistor. a, b,** TDTR phase signal decay curves during (a) reduction and (b) oxidation. The reduction treatment was performed in the order of A, B, C,…F with a step of $Q = 2 \times 10^{21}$ cm$^{-3}$. The oxidation treatment was performed in the order of F, G, H,…K with a step of $Q = 2 \times 10^{21}$ cm$^{-3}$.

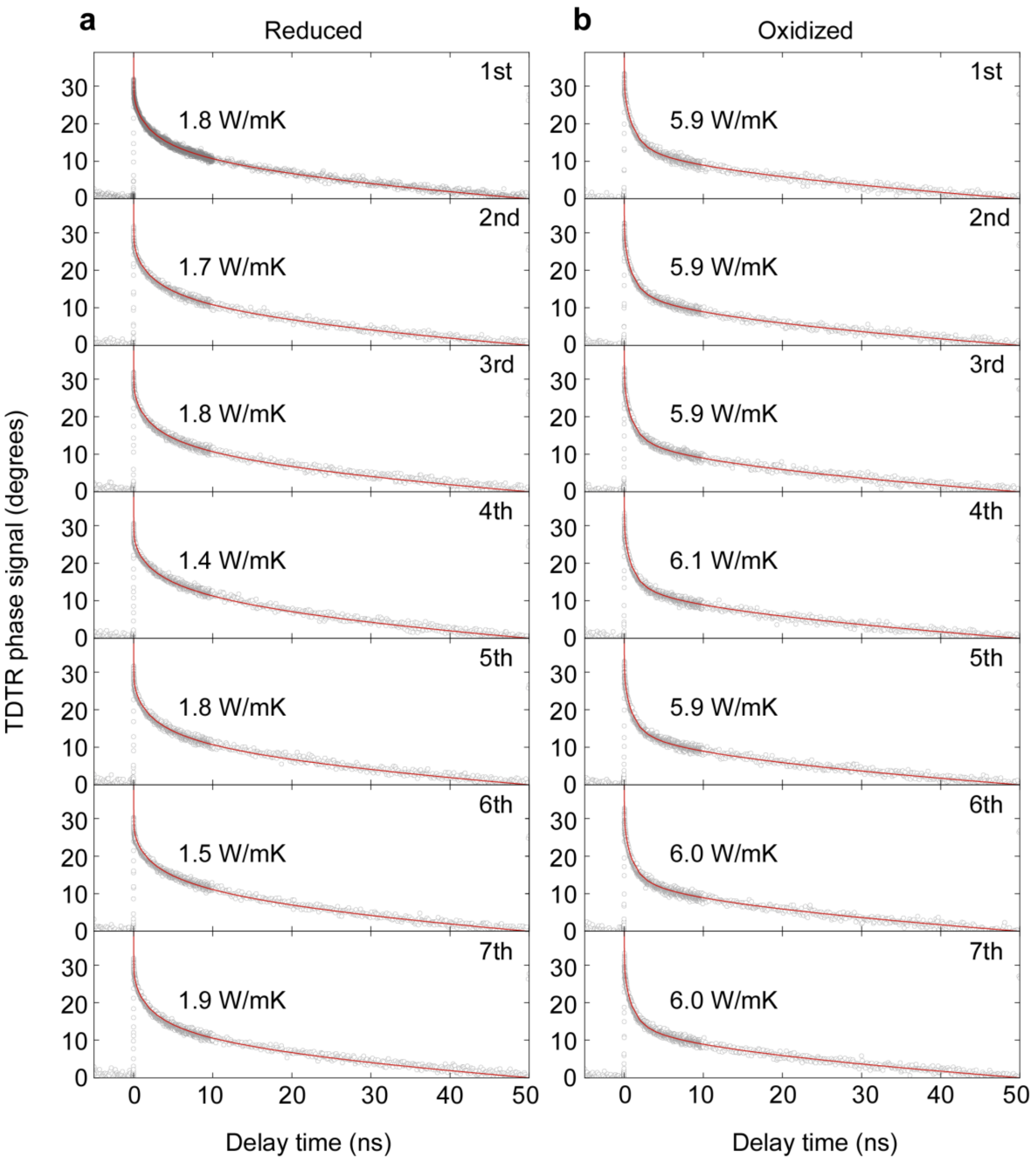

**Figure S4. TDTR decay cycle of the $LaNiO_3$ layer in the thermal transistor. a, b,** TDTR phase signal decay curves after (a) reduction and (b) oxidation.